# Observation of the Isospin-Violating Decay $D_s^{*+} \to D_s^+ \pi^0$

J. Gronberg,[1] C.M. Korte,[1] R. Kutschke,[1] S. Menary,[1] R.J. Morrison,[1] S. Nakanishi,[1] H.N. Nelson,[1] T.K. Nelson,[1] C. Qiao,[1] J.D. Richman,[1] D. Roberts,[1] A. Ryd,[1] H. Tajima,[1] M.S. Witherell,[1] R. Balest,[2] K. Cho,[2] W.T. Ford,[2] M. Lohner,[2] H. Park,[2] P. Rankin,[2] J.G. Smith,[2] J.P. Alexander,[3] C. Bebek,[3] B.E. Berger,[3] K. Berkelman,[3] K. Bloom,[3] T.E. Browder,[3*] D.G. Cassel,[3] H.A. Cho,[3] D.M. Coffman,[3] D.S. Crowcroft,[3] M. Dickson,[3] P.S. Drell,[3] D.J. Dumas,[3] R. Ehrlich,[3] R. Elia,[3] P. Gaidarev,[3] M. Garcia-Sciveres,[3] B. Gittelman,[3] S.W. Gray,[3] D.L. Hartill,[3] B.K. Heltsley,[3] S. Henderson,[3] C.D. Jones,[3] S.L. Jones,[3] J. Kandaswamy,[3] N. Katayama,[3] P.C. Kim,[3] D.L. Kreinick,[3] T. Lee,[3] Y. Liu,[3] G.S. Ludwig,[3] J. Masui,[3] J. Mevissen,[3] N.B. Mistry,[3] C.R. Ng,[3] E. Nordberg,[3] J.R. Patterson,[3] D. Peterson,[3] D. Riley,[3] A. Soffer,[3] P. Avery,[4] A. Freyberger,[4] K. Lingel,[4] C. Prescott,[4] J. Rodriguez,[4] S. Yang,[4] J. Yelton,[4] G. Brandenburg,[5] D. Cinabro,[5] T. Liu,[5] M. Saulnier,[5] R. Wilson,[5] H. Yamamoto,[5] T. Bergfeld,[6] B.I. Eisenstein,[6] J. Ernst,[6] G.E. Gladding,[6] G.D. Gollin,[6] M. Palmer,[6] M. Selen,[6] J.J. Thaler,[6] K.W. Edwards,[7] K.W. McLean,[7] M. Ogg,[7] A. Bellerive,[8] D.I. Britton,[8] E.R.F. Hyatt,[8] R. Janicek,[8] D.B. MacFarlane,[8] P.M. Patel,[8] B. Spaan,[8] A.J. Sadoff,[9] R. Ammar,[10] P. Baringer,[10] A. Bean,[10] D. Besson,[10] D. Coppage,[10] N. Copty,[10] R. Davis,[10] N. Hancock,[10] S. Kotov,[10] I. Kravchenko,[10] N. Kwak,[10] Y. Kubota,[11] M. Lattery,[11] M. Momayezi,[11] J.K. Nelson,[11] S. Patton,[11] R. Poling,[11] V. Savinov,[11] S. Schrenk,[11] R. Wang,[11] M.S. Alam,[12] I.J. Kim,[12] Z. Ling,[12] A.H. Mahmood,[12] J.J. O'Neill,[12] H. Severini,[12] C.R. Sun,[12] F. Wappler,[12] G. Crawford,[13] R. Fulton,[13] D. Fujino,[13] K.K. Gan,[13] K. Honscheid,[13] H. Kagan,[13] R. Kass,[13] J. Lee,[13] M. Sung,[13] C. White,[13] A. Wolf,[13] M.M. Zoeller,[13] X. Fu,[14] B. Nemati,[14] W.R. Ross,[14] P. Skubic,[14] M. Wood,[14] M. Bishai,[15] J. Fast,[15] E. Gerndt,[15] J.W. Hinson,[15] T. Miao,[15] D.H. Miller,[15] M. Modesitt,[15] E.I. Shibata,[15] I.P.J. Shipsey,[15] P.N. Wang,[15] L. Gibbons,[16] S.D. Johnson,[16] Y. Kwon,[16] S. Roberts,[16] E.H. Thorndike,[16] T.E. Coan,[17] J. Dominick,[17] V. Fadeyev,[17] I. Korolkov,[17] M. Lambrecht,[17] S. Sanghera,[17] V. Shelkov,[17] T. Skwarnicki,[17] R. Stroynowski,[17] I. Volobouev,[17] G. Wei,[17] M. Artuso,[18] M. Gao,[18] M. Goldberg,[18] D. He,[18] N. Horwitz,[18] S. Kopp,[18] G.C. Moneti,[18] R. Mountain,[18] F. Muheim,[18] Y. Mukhin,[18] S. Playfer,[18] S. Stone,[18] X. Xing,[18] J. Bartelt,[19] S.E. Csorna,[19] V. Jain,[19] S. Marka,[19] D. Gibaut,[20] K. Kinoshita,[20] P. Pomianowski,[20] B. Barish,[21] M. Chadha,[21] S. Chan,[21] D.F. Cowen,[21] G. Eigen,[21] J.S. Miller,[21] C. O'Grady,[21] J. Urheim,[21] A.J. Weinstein,[21] F. Würthwein,[21] D.M. Asner,[22] M. Athanas,[22] D.W. Bliss,[22] W.S. Brower,[22] G. Masek,[22] and H.P. Paar[22]

(CLEO Collaboration)

[1] *University of California, Santa Barbara, California 93106*
[2] *University of Colorado, Boulder, Colorado 80309-0390*
[3] *Cornell University, Ithaca, New York 14853*
[4] *University of Florida, Gainesville, Florida 32611*
[5] *Harvard University, Cambridge, Massachusetts 02138*
[6] *University of Illinois, Champaign-Urbana, Illinois, 61801*
[7] *Carleton University, Ottawa, Ontario K1S 5B6 and the Institute of Particle Physics, Canada*
[8] *McGill University, Montréal, Québec H3A 2T8 and the Institute of Particle Physics, Canada*
[9] *Ithaca College, Ithaca, New York 14850*
[10] *University of Kansas, Lawrence, Kansas 66045*
[11] *University of Minnesota, Minneapolis, Minnesota 55455*
[12] *State University of New York at Albany, Albany, New York 12222*
[13] *Ohio State University, Columbus, Ohio, 43210*
[14] *University of Oklahoma, Norman, Oklahoma 73019*
[15] *Purdue University, West Lafayette, Indiana 47907*
[16] *University of Rochester, Rochester, New York 14627*
[17] *Southern Methodist University, Dallas, Texas 75275*
[18] *Syracuse University, Syracuse, New York 13244*
[19] *Vanderbilt University, Nashville, Tennessee 37235*
[20] *Virginia Polytechnic Institute and State University, Blacksburg, Virginia, 24061*
[21] *California Institute of Technology, Pasadena, California 91125*
[22] *University of California, San Diego, La Jolla, California 92093*

## Abstract

Using data collected with the CLEO II detector, we have observed the isospin-violating decay $D_s^{*+} \to D_s^+ \pi^0$. The decay rate for this mode, relative to the dominant radiative decay, is found to be $\Gamma(D_s^{*+} \to D_s^+ \pi^0)/\Gamma(D_s^{*+} \to D_s^+ \gamma) = 0.062^{+0.020}_{-0.018} \pm 0.022$.

*Permanent address: University of Hawaii at Manoa

Since the discovery of the charmed strange vector meson $D_s^{*+}$, only the radiative decay mode $D_s^{*+} \to D_s^+\gamma$ has been observed [1,2]. Due to the small mass difference between the $D_s^{*+}$ and the $D_s^+$, the only kinematically allowed strong decay is $D_s^{*+} \to D_s^+\pi^0$. This decay violates strong isospin conservation, since the initial state has $I = 0$ and the final state has $I = 1$. However, isospin is not an exact symmetry, so this decay is not completely forbidden. For example, the decay $\psi' \to J/\psi\,\pi^0$, which has been observed by several experiments [1], violates isospin conservation.

Cho and Wise [3] have made a prediction of the decay rate for $D_s^{*+} \to D_s^+\pi^0$. Using Chiral Perturbation Theory, they describe the decay as an isospin-conserving decay involving a virtual $\eta$, $D_s^{*+} \to D_s^+\eta$; the $\eta$ couples through its $s\bar{s}$ component, so this decay is not suppressed by the Okubo-Zweig-Iizuka rule. This is followed by the $\eta$ mixing into a $\pi^0$. This isospin-violating mixing vanishes in the limit of equal $u$ and $d$ quark masses. The decay amplitude is proportional to the light quark masses in the combination $(m_d - m_u)/[m_s - (m_d + m_u)/2]$, which is $\sim 0.02$–$0.03$ [1]. They also conclude that the radiative decay rate is suppressed (relative to $D^{*0} \to D^0\gamma$) because of the partial cancellation of the magnetic moments of the charm and strange quarks (this type of cancellation also accounts for the small radiative decay rate of the $D^{*+}$ [4]). Thus Cho and Wise relate the rate for $D_s^{*+} \to D_s^+\gamma$ to that for $D^{*+} \to D^+\gamma$ in order to estimate the ratio of partial widths: $R_0 \equiv \Gamma(D_s^{*+} \to D_s^+\pi^0)/\Gamma(D_s^{*+} \to D_s^+\gamma) = \sim 0.01-0.10$. Unfortunately, there are corrections to the prediction which might be negligible, but are presently uncalculable. There is also an electromagnetic amplitude for the decay to $D_s^+\pi^0$, but it is expected [3] to be smaller than the strong amplitude by a factor of order $\alpha/\pi$.

We have searched for $D_s^{*+} \to D_s^+\pi^0$ using the decay chain $D_s^+ \to \phi\pi^+$, $\phi \to K^+K^-$. At the same time, we observe the radiative decay to normalize the hadronic decay rate. Since the same charged-track selection criteria are used for both decay modes, only the relative efficiencies for finding a single photon or reconstructing a $\pi^0$ are needed.

The data used in this analysis were collected with the CLEO II detector at CESR. The detector consists of a charged particle tracking system surrounded by time-of-flight (TOF) scintillation counters. These are in turn surrounded by an electromagnetic calorimeter which consists of 7800 thallium-doped CsI crystals. The inner detector is immersed in a 1.5 T solenoidal magnetic field generated by a superconducting coil. Finally, the magnet coil is surrounded by iron flux return and muon counters. Charged particle identification is provided by specific ionization measurements in the main drift chamber and by TOF measurements. A detailed description of the detector can be found elsewhere [5].

The data were taken at center-of-mass energies equal to the masses of the $\Upsilon(3S)$ and $\Upsilon(4S)$, and in the continuum above and below the $\Upsilon(4S)$. The total integrated luminosity is 3.75 fb$^{-1}$. Events were required to have a minimum of three charged tracks, and energy in the calorimeter greater than 15% of the center-of-mass energy. Charged tracks were initially required to pass a loose particle identification consistency. We required that the specific ionization measurement be within three standard deviations of that expected for the hypothesis in question, either kaon or pion.

Only energy clusters in the barrel calorimeter with $|\cos\theta| \leq 0.71$ (where $\theta$ is the polar angle with respect to the beamline) which were not matched to tracks were used as photons. They were required to have a minimum of energy of 30 MeV, and to pass a lateral shape cut to help eliminate energy from hadronic interactions. Single photons used to reconstruct the radiative decay were required to have energy greater than 50 MeV. Pairs of photons were used to reconstruct $\pi^0$'s. The invariant mass of the two photons was required to be within 2.5 standard deviations of the $\pi^0$ mass; this cut takes into account the asymmetric $\pi^0$ lineshape and the small momentum dependence of the mass resolution. The $\pi^0$ candidates were kinematically fit to the $\pi^0$ mass to improve momentum resolution. The decay angle, $\theta_\gamma$, is defined as the angle between the direction of one of the photons in the $\pi^0$ rest-frame and the $\pi^0$ direction in the lab-frame. We required $|\cos\theta_\gamma| \leq 0.75$, since the background peaks near $|\cos\theta_\gamma| = 1$ while the signal is flat.

We began the reconstruction by taking pairs of oppositely charged tracks, consistent with being kaons, and calculating the invariant mass. Those pairs whose invariant mass was within $\pm 9$ MeV/$c^2$ of the $\phi$ mass [1] were accepted as $\phi$ candidates. Each remaining charged track, consistent with being a pion, was combined with the $\phi$ to make a $D_s^+$ candidate. The $D_s^+$ candidates were required to pass two angle cuts. To reduce background from slow pions, we required $\cos\theta_\pi \geq -0.9$, where $\theta_\pi$ is the decay angle of the $\pi^+$ (the angle between the pion's direction in the $D_s^+$ rest frame, and the $D_s^+$'s direction in the lab frame). The signal distribution is flat, while the background peaks near $\cos\theta_\pi = -1$. Second, because the $\phi$ is polarized in the helicity-zero state, the kaons must have a helicity angle distribution proportional to $\cos^2\theta_K$, where $\theta_K$ is the angle between the kaon and the $D_s^+$, both measured in the $\phi$ rest-frame. We required $|\cos\theta_K| \geq 0.35$.

Next, we imposed a more restrictive particle identification cut on the three-track combination. A particle ID $\chi^2$ was calculated using the specific ionization measurements for each of the three tracks and the TOF measurements for each track which had good TOF information. We required that the $\chi^2$ probability be at least 0.1. Finally, $D_s^+$ candidates had to have a mass within two standard deviations ($\pm 16$ MeV/$c^2$) of the $D_s^+$ mass [1].

To reconstruct the radiative decay mode, the $D_s^+$ candidates were combined with each photon in the event which had an energy of at least 50 MeV. The scaled momentum, $x$, of each $D_s^{*+}$ candidate was calculated as $x = p/p_M$, where $p_M^2 = E_0^2 - M_{D_s^{*+}}^2$, and $E_0$ is the beam energy; we required $x \geq 0.6$. The mass difference, $\Delta M_\gamma \equiv M(D_s^+\gamma) - M(D_s^+)$ was calculated and histogrammed [6]. The resulting distribution was fit using a Gaussian modified with an enhanced low-energy tail for the signal and a third-order polynomial for the background. We find $944 \pm 57$ signal events (statistical error only).

Next, each $D_s^+$ candidate was combined with each $\pi^0$ with momentum of at least 250 MeV/$c$. An $x$ cut of $x \geq 0.6$ was again applied to each $D_s^{*+}$ candidate. In Fig. 1, we show the mass difference, $\Delta M_\pi \equiv M(D_s^+\pi^0) - M(D_s^+)$ for the remaining $D_s^{*+}$ candidates. The data were fit using a Gaussian for the signal and a square-root function that goes to zero at threshold to represent the background. The r.m.s. width of the Gaussian was fixed at $\sigma = 1.2$ MeV/$c^2$, as determined by the Monte Carlo. The mean was fixed at 144.22 MeV/$c^2$, as previously measured by CLEO [7]. We find $14.7^{+4.6}_{-4.0}$ signal events (statistical errors only).

We also studied the $D_s^+$ and $\pi^0$ sidebands. We used $\phi\pi^+$ combinations with masses between 1904 and 1936 MeV/$c^2$, and between 2004 and 2036 MeV/$c^2$, and combined them with $\pi^0$'s. We also selected $\gamma\gamma$ combinations that were between 2.75 and 7.75 standard deviations away from the $\pi^0$ mass; this corresponds approximately to an invariant mass between 88 and 118 MeV/$c^2$, or between 145 and 165 MeV/$c^2$; these were combined with the $D_s^+$ candidates. These two sets of sideband combinations produce the $\Delta M_\pi$ distribution shown as a dashed histogram in Fig. 1. The entries in this histogram have been scaled by a factor of 0.5, to account for the fact that the sidebands are twice as wide as the signal band. If we fit this sideband histogram the same way as the signal band events, the area of the Gaussian is $-1.0^{+3.1}_{-2.4}$ events, consistent with zero.

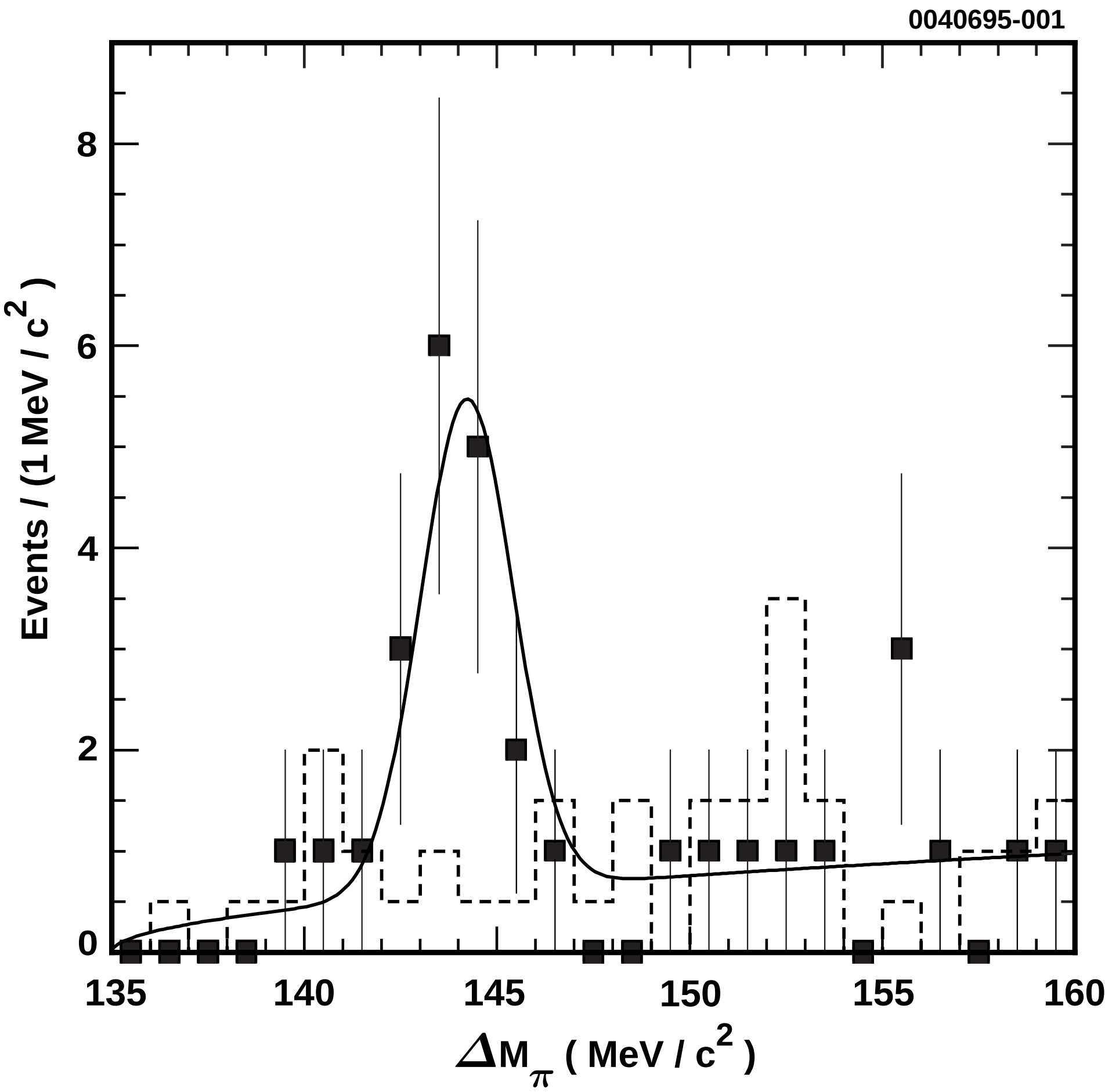

FIG. 1. The distribution of the mass difference, $\Delta M_\pi$, for the isospin-violating decay $D_s^{*+} \to D_s^+\pi^0$. The points are the data after all cuts, the solid line is the fit to the data, and the dashed line is an estimate of the backgrounds, as described in the text.

As a check, we have also analyzed events in the Cabibbo-suppressed decay chain $D^{*+} \to D^+\pi^0$, $D^+ \to \phi\pi^+$. We used the same cuts as before, but required the $\phi\pi^+$ invariant mass to lie in the $D^+$ signal region from 1853.3 to 1885.3 MeV/$c^2$. Using a Gaussian with mean fixed at 140.64 MeV/$c^2$, as measured by CLEO [8], we find $28.6^{+6.0}_{-5.4}$ signal events (statistical errors only). Using our measured luminosity, and published cross sections [9] and branching ratios [1], we expect $23.7 \pm 4.2$ events.

In order to confirm that this signal is from $D_s^{*+} \to D_s^+\pi^0$ rather than $D_s^{*+} \to D_s^+\gamma\gamma$ we have relaxed the cut on the two photon invariant mass used in the $\pi^0$ selection. We then selected events with 141.22 MeV/$c^2 \leq \Delta M_\pi <$ 147.22 MeV/$c^2$, and studied the $\gamma\gamma$ invariant mass distribution. Fitting this distribution yields a $\pi^0$ signal of $16.1^{+4.6}_{-4.0}$ events, consistent with our previous result. Similarly, when we cut on $\Delta M_\pi$ and fit the $\phi\pi^+$ invariant mass distribution we find $13.9^{+4.8}_{-4.1}$ signal events; this is shown in Fig. 2.

The background appears to be dominated by random combinations, rather than feed-through from some other physics channel. For example, two other conceivable sources of background, real $D_s^{*+} \to D_s^+ \gamma$ events with an extra soft photon faking a $\pi^0$, and misidentified $D^{*+} \to D^+ \pi^0$ events are both negligible. We have generated $D_s^{*+} \to D_s^+ \gamma$ Monte Carlo events and analyzed them with the reconstruction program. The Monte Carlo sample is 50% larger than our actual data sample. Only 3 events with 135 MeV/$c^2 \leq \Delta M_\pi <$ 160 MeV/$c^2$ are found; they are all outside the signal region 141 MeV/$c^2 \leq \Delta M_\pi <$ 147 MeV/$c^2$.

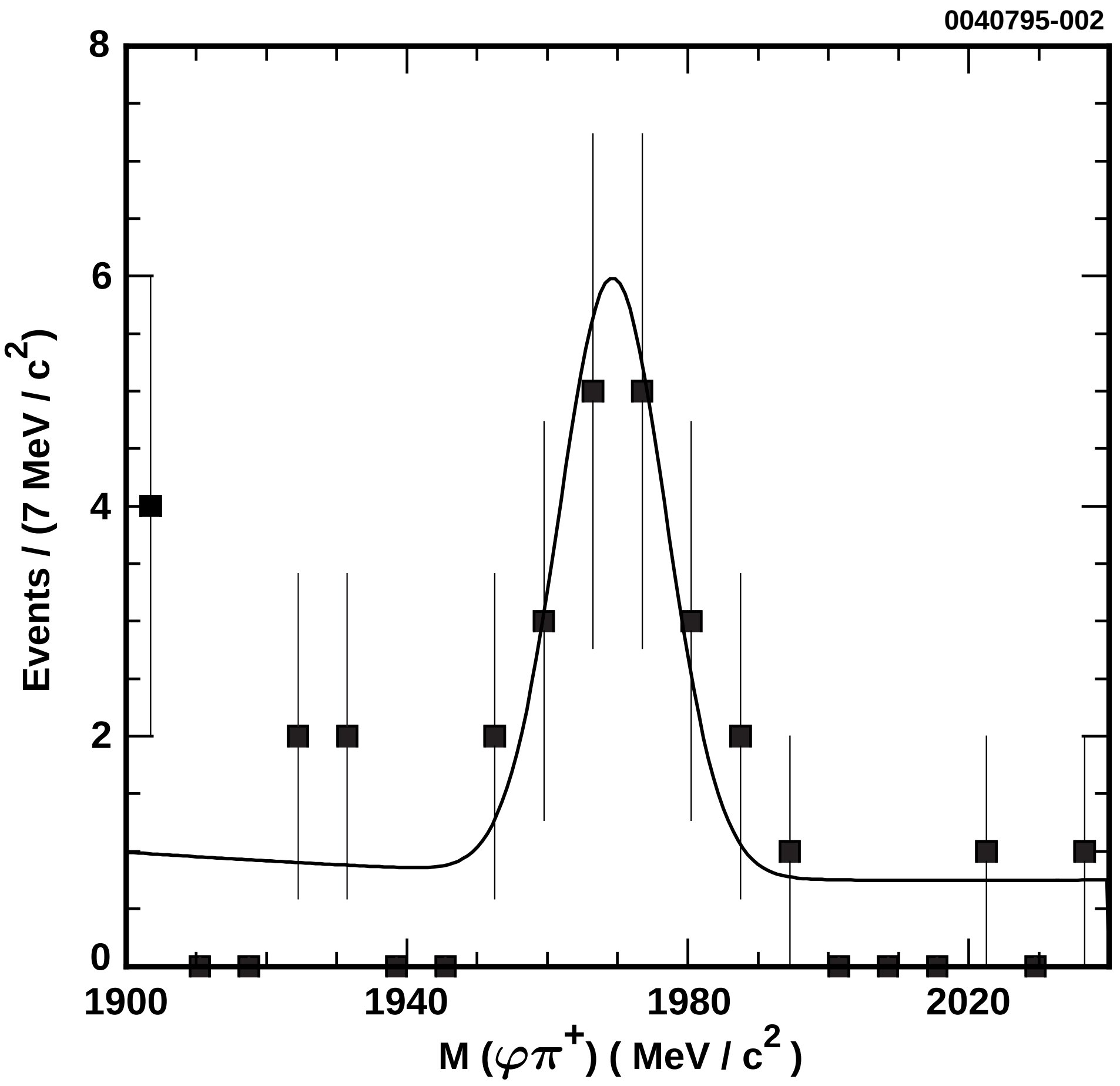

FIG. 2. The $M(\phi\pi^+)$ distribution for events in the $\Delta M_\pi$ signal region. The points are the data after all cuts; the solid line is the fit to the data, using a Gaussian of fixed mean and width for the signal, and a second-order polynomial for the background.

Similarly, background from misidentifed $D^{*+}$ decays is not a problem. The particular decay chain considered is: $D^{*+} \to D^+\pi^0$, $D^+ \to K^-\pi^+\pi^+$, where one of the pions is misidentified as a kaon, so that the three charged tracks reconstruct near the $D_s^+$ mass. These events are not a problem because it is almost impossible for such a fake kaon to make a $\phi$, and then for the fake $\phi$ plus the real pion to make a $D_s^+$. This was tested by taking each $D_s^+$ candidate and changing the particle identification of the kaon with the same sign as the pion, from kaon to pion. We recalculated the invariant mass of the the three tracks and found that none of the calculated masses were greater than 1845 MeV/$c^2$. Studies of the data and Monte Carlo events confirmed that such $D^{*+}$ events could not produce this peak. Other similar decays of charmed mesons, which are partially reconstructed, and include misidentified particles, can contribute to the signal histogram. Monte Carlo studies showed, however, that such events do not form a peak.

We have evaluated the statistical significance of $D_s^{*+} \to D_s^+ \pi^0$ events in two ways. First, we refit the signal histogram (Fig. 1), constraining the area of the Gaussian to zero. The change in likelihood from the original fit is equivalent to 5.0 standard deviations.

Second, we counted the number of events with 142 MeV/$c^2 \leq \Delta M_\pi <$ 146 MeV/$c^2$ in both the signal histogram and the sideband histogram. This yields 16 signal events and 5 sideband events. Taking into account that the sideband width is twice the signal band width, the binomial probability of getting 16 (or more) signal events out of 21 total events is $7.3 \times 10^{-5}$, equivalent to 3.9 standard deviations. Thus we consider the observation to have at least 3.9 standard deviation significance.

Using the CLEO Monte Carlo program, we found that the ratio of efficiencies for reconstructing the hadronic and radiative events is $0.25 \pm 0.03$. Using this ratio and the numbers of events reconstructed, we find that the ratio of partial widths is $R_0 = 0.062^{+0.020}_{-0.018}$, where these statistical errors are dominated by the error on the number of $D_s^+ \pi^0$ events.

We estimate that the systematic error on $R_0$ is 35%. This is dominated by variations in $R_0$ when we vary our cuts on the $\pi^0$ and/or photons. Thus our measurement of the ratio of partial widths is [10]:

$$R_0 = \frac{\Gamma(D_s^{*+} \to D_s^+ \pi^0)}{\Gamma(D_s^{*+} \to D_s^+ \gamma)} = 0.062^{+0.020}_{-0.018} \pm 0.22. \tag{1}$$

If we assume that these two branching fractions sum to one, the individual branching fractions are $\mathcal{B}(D_s^{*+} \to D_s^+ \pi^0) = 0.058^{+0.018}_{-0.016} \pm 0.020$ and $\mathcal{B}(D_s^{*+} \to D_s^+ \gamma) = 0.942^{+0.016}_{-0.018} \pm 0.020$.

The observation of this decay mode implies that the $D_s^{*+}$ must have natural spin-parity ($0^+$, $1^-$, $2^+ \ldots$), since conservation of parity and angular momentum forbid the decay of a particle with unnatural spin-parity to two pseudoscalars. The radiative decay rules out $0^\pm$; the most likely spin-parity is $J^P = 1^-$, the same as the $D^{*0}$ and $D^{*+}$ [1].

Using the $D_s^{*+} \to D_s^+ \pi^0$ events, we can also make a measurement of the mass difference $M_{D_s^{*+}} - M_{D_s^+}$, and set an upper limit on the width of the $D_s^{*+}$. We refit the $\Delta M_\pi$ distribution, allowing the mean of the signal Gaussian to float. With the width of the Gaussian fixed, the mean is fit to $\Delta M_\pi = 143.76 \pm 0.39$ MeV/$c^2$ (statistical error only). Fitting the $D^{*+} \to D^+ \pi^0$, $D^+ \to \phi \pi^+$ events mentioned above, we find $M_{D^{*+}} - M_{D^+} = 140.31 \pm 0.26$ MeV/$c^2$, in good agreement with the previous CLEO measurement of $140.64 \pm 0.10$ MeV/$c^2$ [8]. We include the 0.33 MeV/$c^2$ difference in the systematic error of the $\Delta M_\pi$ measurement. It has previously been estimated that the uncertainty in the crystal energy calibration introduces a systematic error of 0.04 MeV/$c^2$ in this type of measurement [8]. Changing the cuts used to select the events introduces variations of 0.22 MeV/$c^2$; other effects, such as varying the background function used for the fit, produce much smaller variations. Thus we estimate the systematic error to be 0.40 MeV/$c^2$. Therefore we measure $M_{D_s^{*+}} - M_{D_s^+} = 143.76 \pm 0.39 \pm 0.40$ MeV/$c^2$, in excellent agreement with the previous CLEO value $144.22 \pm 0.47 \pm 0.37$ MeV/$c^2$ from the radiative mode [7]. The two measurements are statistically independent, and have almost completely independent systematic errors. Averaging the two, we find $M_{D_s^{*+}} - M_{D_s^+} = 143.97 \pm 0.41$ MeV/$c^2$. This may be compared with the Particle Data Group's fit value of $141.6 \pm 1.8$ MeV/$c^2$, or their average value of $142.4 \pm 1.7$ MeV/$c^2$ [1]; the PDG's values do not include the earlier CLEO measurement.

The width of the signal is consistent with being entirely due to detector resolution. The Monte Carlo calculation predicts a value of $\sigma = 1.19 \pm 0.07$ MeV/$c^2$ for the signal Gaussian (statistical error only). The measured r.m.s. width of the signal Gaussian is $\sigma = 1.06^{+0.41}_{-0.28}$ MeV/$c^2$. Assuming a 10% systematic error on the Monte Carlo prediction for $\sigma$, and fitting the signal with a $p$-wave Breit-Wigner convoluted with a Gaussian, we can set a 90% confidence level upper limit $\Gamma(D_s^{*+}) < 1.9$ MeV/$c^2$. This can be compared with the best existing limit, from ARGUS [11], $\Gamma(D_s^{*+}) < 4.5$ MeV/$c^2$.

In conclusion, we have detected the isospin-violating decay $D_s^{*+} \to D_s^+ \pi^0$, and find the ratio of partial widths, $R_0 = 0.062^{+0.20}_{-0.18} \pm 0.022$, which confirms a recent prediction [3,12]. We determine the branching fractions for the $D_s^+\gamma$ and $D_s^+\pi^0$ modes assuming that any other decay modes of the $D_s^{*+}$ are negligible. In addition, we present a new measurement of $M_{D_s^{*+}} - M_{D_s^+}$, and we obtain an improved upper limit on the width of the $D_s^{*+}$. The observation of this decay mode implies that the $D_s^{*+}$ has natural spin-parity, most likely $1^-$.

One of us (J.B.) would like to thank Peter Cho and Mark Wise for stimulating discussions. We gratefully acknowledge the effort of the CESR staff in providing us with excellent luminosity and running conditions. This work was supported by the National Science Foundation, the U.S. Department of Energy, the Heisenberg Foundation, the Alexander von Humboldt Stiftung, the Natural Sciences and Engineering Research Council of Canada, and the A.P. Sloan Foundation.